\documentclass[12pt]{article}
\usepackage{amssymb,amsmath,epsfig}
\begin{document}

\title{\bf Magnetohydrodynamic Viscous Flow
Over a Shrinking Sheet With Second Order Slip Flow Model}

\author{Tahir Mahmood$^a$ and S. Munawar
Shah$^a$, G. Abbas$^b$\thanks{ghulamabbas@ciitsahiwal.edu.pk},
\\\\
$^a$Department of Mathematics, The Islamia University\\
Bahawalpur, Pakistan\\
$^b$Department of Mathematics, COMSATS Institute\\
of
Information Technology, Sahiwal, Pakistan.}

\date{}
\maketitle
\begin{abstract}
In this paper, we investigate the magnetohydrodynamic viscous flow
with second order slip flow model over a permeable shrinking
surface. We have obtained the closed form of exact solution of
Navier-Stokes equations by using similarity variable technique. The
effects of slip, suction and magnetic parameter have been
investigated in detail. The results show that there are two solution
branches, namely lower and upper solution branch. The behavior of
velocity and shear stress profiles for different values of slip,
suction and magnetic parameters has been discussed through graphs.
\end{abstract}

\section{Introduction}

The fluid flow due to a moving boundary has a great important in the
expulsion processes \cite{1}-\cite{3}. The boundary layer on a
stretching surface with a constant speed has been studied first time
by Sakiadis \cite{4,5}. Later on Wang \cite{6} pointed out that
solution presented by Sakiadis was not an exact solution of
Navier-Stokes solutions. The exact solution of two dimensional
Navier-Stokes equations over a stretching sheet sheet was presented
by Crane \cite{7}. Gupta and Gupta \cite{8}, have examined the
effect of mass transfer on Crane flow. For three dimensional
background, the stretching sheet problem was investigated by Wang
\cite{9}. In this paper, the exact similarity solutions of
Navier-Stokes equations were obtained. This study was extended to
the rotating flow over double stretchable disks \cite{10}. All these
solutions are exact solutions of the Navier-Stokes equations.

The magnetohydrodynamic (MHD) flow over a stretching sheet was
studied extensively \cite{11}-\cite{14} in the recent years, for
both impermeable surface and permeable surface. The micro-scale
fluid dynamics in the Micro-Electro-Mechanical Systems has received
a considerable interest in among the researchers. The flow of such
fluids belongs to the slip flow regime, which is entirely different
from the traditional flow \cite{15}. For the flow in the slip
regime, the fluid flow obeys the Navier–Stokes equations in the
presence of slip velocity or temperature boundary conditions. In
addition, partial velocity slips over a moving surface occur for
particular fluids such as emulsions, suspensions and polymer
solutions \cite{16}.

The slip flows under different flow configurations have been
investigated by various authors \cite{17}-\cite{22}. With a slip at
the wall boundary, the flow behavior with slip conditions are
different from no-slip conditions. Among these studied, Andersson
\cite{17} and \cite{18} investigated the slip flow over an
impermeable stretching surface. In a recent paper, Wang \cite{22}
examined the mass transfer effects on the slip flow over a
stretching surface. However, there was no study on the slip MHD flow
over a permeable stretching surface in the literature. Fang et
al.\cite{23} studied study the first order slip MHD flow over a
permeable stretching surface. Exact solutions of the governing
Navier-Stokes equations were presented.  Also, Fung et al.\cite{24}
studied the viscous flow over a shrinking sheet with second order
slip flow model. However, there was no study on MHD viscous flow
over a shrinking sheet with second order slip flow model. Therefore,
the objective of the current paper is to study the second order slip
MHD flow over a permeable shrinking surface. Exact solutions of the
governing Navier-Stokes equations are presented and discussed in
detail.
\section{Mathematical Formulation and Solution of
Problem}

We consider two dimensional laminar flow of magnetized viscous fluid
over a continuously shrinking sheet. The shrinking velocity of the
sheet is $U_w =-U_0x$, where $U_0$ is a constant and $v_w = v_w(x)$
is wall mass transfer velocity. The x-axis is along the shrinking
surface in the direction opposite the sheet motion and y-axis is
perpendicular to it. By these assumptions the corresponding
Navier-Stoke's equations for the present problem can be summarized
by the following set of equations \cite{13}-\cite{15}, \cite{19,20}:
\begin{eqnarray}\label{1}
&&\frac{\partial u} {\partial x} + \frac{\partial u} {\partial y} =
0;\\\label{2} &&u \frac{\partial u} {\partial x} + v \frac{\partial
u} {\partial y} = -\frac{1}{\rho}
 \frac {\partial p}{\partial x} + \nu( \frac{\partial^2u} {\partial x^2}
+\frac{\partial^2 u} {\partial y^2} ) - \frac{\sigma B^2u}{\rho},
\\\label{3}
&&u\frac{\partial v} {\partial x} + v \frac{\partial v} {\partial y}
= -\frac{1}{\rho}\frac{\partial p}{\partial x} + \nu(
\frac{\partial^2v} {\partial x^2} + \frac{\partial^2 v} {\partial
y^2}).
 \end{eqnarray}
The boundary conditions \cite{24} are
\begin{equation}\label{4}
u(x,0)=U_0(x) + U_{slip}, v(x, 0)=v_w, u(x,1)=0.
\end{equation}
where $u$ and $v$ are the components of velocity in the $x$ and $y$
directions respectively, $\nu$ is the kinematic viscosity, $p$ is
the fluid pressure, $\rho$ is fluid density and $U_{slip}$ is second
order velocity slip which is valid for any arbitrary Knudsen
numbers, $K_n$ \cite{25} and is given by
\begin{eqnarray}\nonumber
U_{slip} &=& \frac{2}{3}\left(\frac{3-\alpha
l^3}{\alpha}+\frac{3}{2}(\frac{1-\dot{l}^2}{K_n})\right)\lambda
\frac{\partial u}{\partial y}-\frac{1}{4}\left(l^4 + \frac{2}
{{K_n}^2} (1 -l^2)\right){\lambda}^2 \frac{\partial ^2 u}{\partial
y^2}\\\label{5} &=&A\frac{\partial u}{\partial y}+B \frac{\partial
^2 u}{\partial y^2}
\end{eqnarray}
where $l = min\left[\frac{1}{K_n}-1\right]$, $0\leq\alpha\leq1$ is
the momentum accommodation coefficient and $\lambda>0$ is the
molecular mean free path. According to the definition of $l$, it is
observed that $0\leq l\leq1$, for any value of $K_n$. The stream
function and similarity variable would be of the following form
\begin{equation}\label{6}
\psi(x,y) =f(\eta)x \sqrt{vU_0},~~~\eta(x,y)= y\sqrt{\frac{U_0} {v}
}.
\end{equation}
With these definitions the components of velocities are
\begin{equation}\label{7}
u=f'(\eta)x U_0,~~v=-f(\eta)\sqrt{vU_0}.
\end{equation}
The wall mass transfer velocity becomes
\begin{equation}\label{8}
v_w =f(0)\sqrt{vU_0}.
\end{equation}
Using Eqs.(\ref{7}) and (\ref{8}) in (\ref{3})
\begin{equation}\label{9}
\frac{p}{\rho}= \nu \frac{\partial v} {\partial y}- \frac{v^2}{2} +
constant
\end{equation}
Also, by using Eqs.(\ref{7})-(\ref{9}) in Eq.(\ref{2}), we obtain
\begin{equation}\label{10}
f'''+ ff'' - {f'}^2-M^2f' = 0,
\end{equation}
where $M^2=\frac{\sigma B^2}{\rho \mu_0}$. The boundary conditions
are
\begin{eqnarray}\label{11}
&&f'(0) =-1 + \gamma f''(0) +\delta f'''(0), \\\label{12} &&f(0)=
s\\\label{13} &&f'(1) = 0
\end{eqnarray}
where $s$ is the mass transfer parameter, $\gamma$ is the first
order velocity slip parameter with $\gamma =
A\sqrt{\frac{U_0}{\nu}}$ and $\delta =\frac{BU_0}{\nu}< 0$ is the
second order velocity slip parameter. The pressure term can be
obtained from Eq.(9). In this section, we are interested to find the
closed form of exact solution of Eq.(\ref{10}) subjected to boundary
conditions (\ref{11})-(\ref{13}).

We assume the solution of the form
\begin{equation}\label{14}
f(\eta) = a + be^{-\beta\eta}
\end{equation}
Using above equation in the boundary condition (\ref{12}), we get
\begin{equation}\label{15}
 s= a + b.
\end{equation}
Equation (\ref{11}) leads to
\begin{equation}\label{16}
 b= \frac{1}{\beta+\gamma{\beta}^2-\delta {\beta}^3}.
\end{equation}
From Eqs.(\ref{15}) and (\ref{16}), we get
\begin{equation}\label{17}
 a= s-\frac{1}{\beta+\gamma{\beta}^2-\delta {\beta}^3}.
\end{equation}

Using the values of $a$ and $b$ in the assumed solution (\ref{14}),
we get
\begin{equation}\label{18}
f(\eta) = s-\frac{1}{\beta+\gamma{\beta}^2-\delta
{\beta}^3}+\frac{1}{\beta+\gamma{\beta}^2-\delta
{\beta}^3}e^{-\beta\eta}.
\end{equation}
The use of this relationship in Eq.(\ref{10}), leads to the
following fourth order algebraic equation for $\beta$
\begin{equation}\label{19}
\beta^4 + A_3{\beta}^3 + A_2\beta^2 + A_1\beta + A_0 = 0;
\end{equation}
where
\begin{equation}\label{20}
A_3 =\frac{-\gamma-\delta s}{\delta},~~A_2 =\frac{\gamma s-\delta
M^2-1} {\delta},~~ A_1 = \frac{s+\gamma M^2}{\delta},~~A_0 = \frac{
M^2-1}{\delta}.
\end{equation}
Now our focus is to find the solution of above equation, such
solution can be determined by wall mass transfer $s$, first order
slip parameter $\gamma$ and second order slip parameter $\delta$ .
The only positive roots of Eq.(\ref{19}) are solutions. By defining
$z=\beta+\frac{A_3}{4}$, Eq.(\ref{19}) can be converted to following
standard form
\begin{equation}\label{21}
z^4 + P{z}^2 + qz +r = 0,
\end{equation}
where
\begin{eqnarray}\nonumber
&&P=A_2+\frac{3}{8}{A_3}^2,~~q=A_1 - \frac{1} {2} A_2A_3 + \frac{1}
{8}{A_3}^3,\\\label{22} && r=A_0-\frac{1}{4} A_1{A_3}+ \frac{1}{16}
A_2{A_3}^2-\frac{3}{256}{A_3}^4.
\end{eqnarray}
Mathematically, there are four roots of Eq.(\ref{21}), real roots of
this equation can be written explicitly in the following form
\begin{eqnarray}\nonumber
&&\beta_1=\frac{\sqrt{c}}{2}-\frac{1}{2}\sqrt{D-\frac{2q}{\sqrt{c}}}-A_3\\\label{23}
&&\beta_2=\frac{\sqrt{c}}{2}+\frac{1}{2}\sqrt{D-\frac{2q}{\sqrt{c}}}-A_3,
\end{eqnarray}
where
\begin{eqnarray}\nonumber
&c=&\frac{-2P}{3}+\frac{2^{\frac{1}{3}}(P^2+12r)}{\left(3(2P^2 +
27q^2- 72Pr+\sqrt{-4(P^2 + 12r)^3 + (2P^2 + 27q^2-72Pr)^2
)}^{\frac{1}{3}}\right)}\nonumber\\&+& \frac{\left(3\left(2P^2 +
27q^2- 72Pr+\sqrt{-4(P^2 + 12r)^3 + (2P^2 + 27q^2-72Pr)^2
}\right)^{\frac{1}{3}}\right)}{3\times 2^{\frac{1}{3}}}\nonumber\\
\\
&D=&\frac{-4P}{3}-\frac{2^{\frac{1}{3}}(P^2+12r)}{\left(3(2P^2 +
27q^2- 72Pr+\sqrt{-4(P^2 + 12r)^3 + (2P^2 + 27q^2-72Pr)^2
)}^{\frac{1}{3}}\right)}\nonumber\\&+& \frac{\left(3\left(2P^2 +
27q^2- 72Pr+\sqrt{-4(P^2 + 12r)^3 + (2P^2 + 27q^2-72Pr)^2
}\right)^{\frac{1}{3}}\right)}{3\times 2^{\frac{1}{3}}}\nonumber\\
\end{eqnarray}
Here $\beta_1$ and $\beta_2$ corresponds to lower and upper branch
solution, respectively.
\begin{figure}
\center\epsfig{file=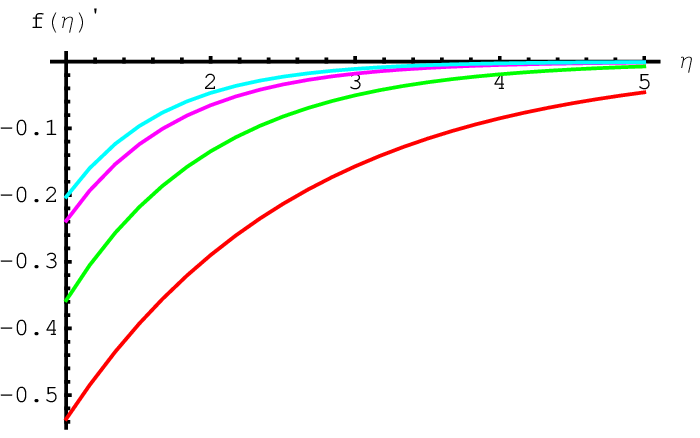, width=0.45\linewidth} \epsfig{file=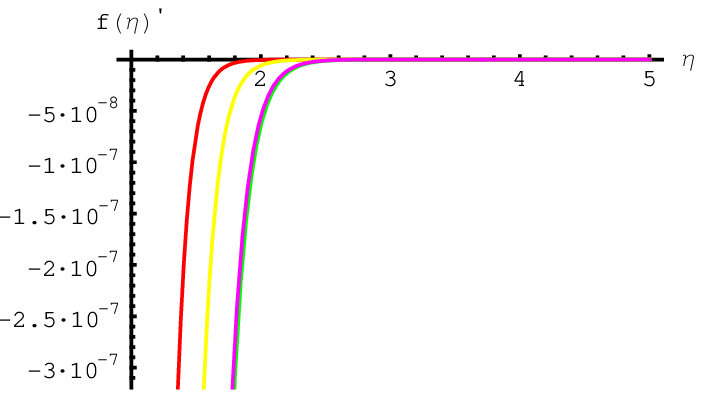,
width=0.45\linewidth}\caption{The left and right figures describe
the behavior of $f'(\eta)$ corresponding to lower and upper branch
solution, respectively for varying $\gamma$ and $s=1,~\delta=-0.01,~
M=0.1.$ In the left figure light green, pink, green and red curves
corresponds to $\gamma= 0.08,~ 0.04,~0.01,~0.1,$ respectively while
in right figure red, brown, pink and green curves correspond to
$\gamma= 0.08,~ 0.04,~0.01,~0.1,$ respectively.}
\end{figure}

To show the effects of slip and magnetic parameters on the fluid
flow, we present the velocity and shear stress profiles for
different combinations of $M,~\delta,~\gamma$ and $s$. The velocity
profile is presented in ¯figures \textbf{1-4}. The left graph in
figure \textbf{1} shows that for lower branch solution the velocity
of the shrinking sheet increases with the increase of $\gamma$ by
keeping all the other parameters fixed. The right graph in figure
\textbf{1} shows that for upper branch solution the velocity of the
shrinking sheet increases from $-\infty$ value to zero, with the
increase of $\gamma$. The similar behavior of velocity profile is
depicted in the both graphs of figure \textbf{2} for lower and upper
branch solutions, respectively, with varying $s$.
\begin{figure}
\center\epsfig{file=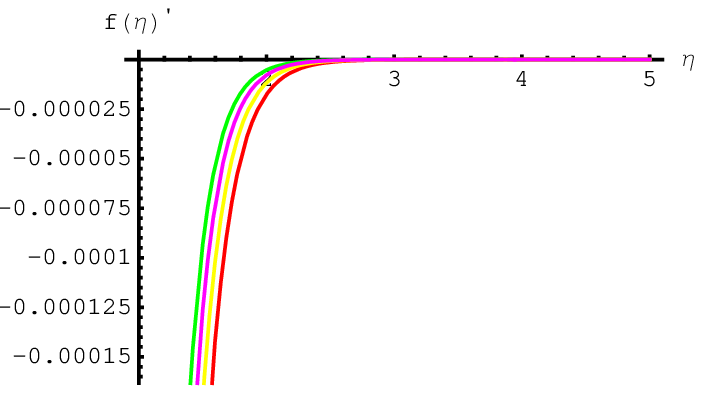, width=0.45\linewidth} \epsfig{file=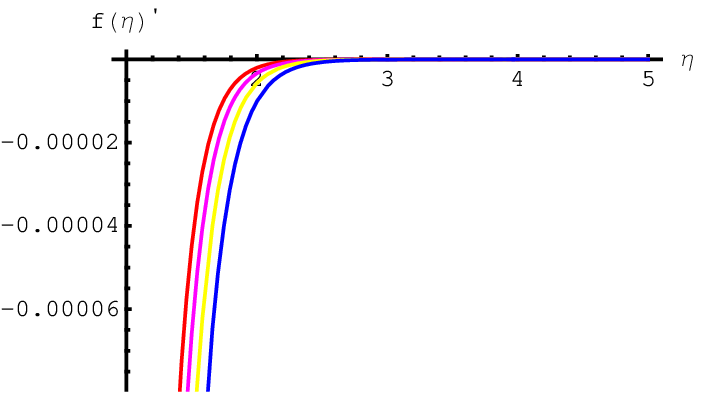,
width=0.45\linewidth}\caption{The left and right figures describe
the behavior of $f'(\eta)$ corresponding to lower and upper branch
solution, respectively for varying $s$ and $\gamma= 0.1,~\delta =
-0.01,~M=0.1.$ In the left figure red, brown, pink and green curves
corresponds to $s=0.1,~0.2,~0.3,~0.4,$ respectively while in right
figure blue, brown, pink and red curves correspond to
$s=0.1,~0.2,~0.3,~0.4,$respectively.}
\end{figure}
\begin{figure}
\center\epsfig{file=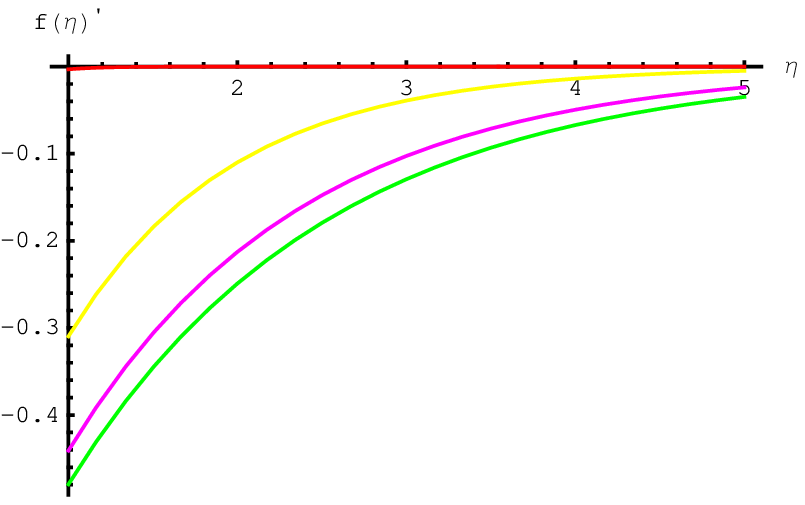, width=0.45\linewidth} \epsfig{file=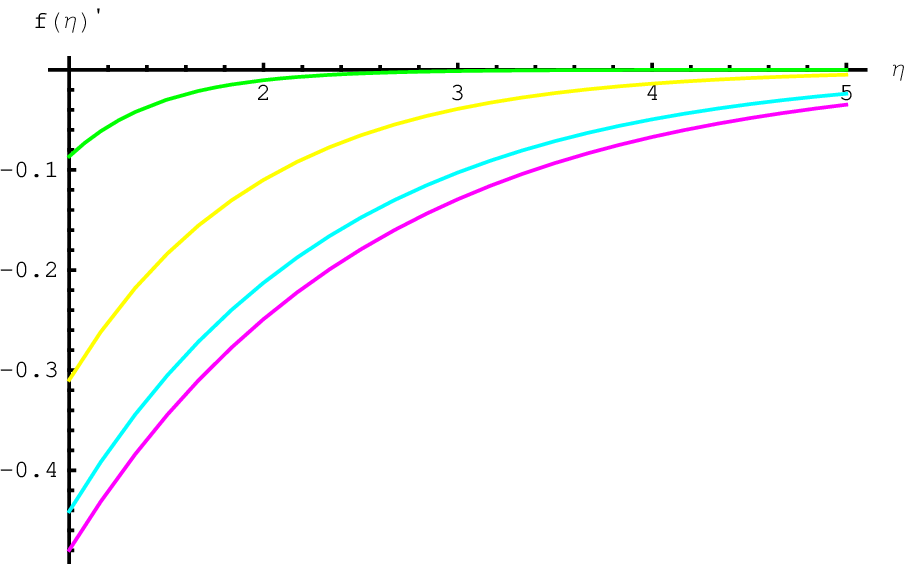,
width=0.45\linewidth}\caption{The left and right figures describe
the behavior of $f'(\eta)$ corresponding to lower and upper branch
solutions, respectively for varying $M$ and $\gamma=0.1,~\delta=
-0.01,~s = 1.$ In the left figure red, brown, pink and green curves
corresponds to $M =0.1,~2,~3,~4,$ respectively while in right figure
green, brown, light green and pink curves correspond to $M
=0.1,~2,~3,~4,$ respectively.}
\end{figure}
\begin{figure}
\center\epsfig{file=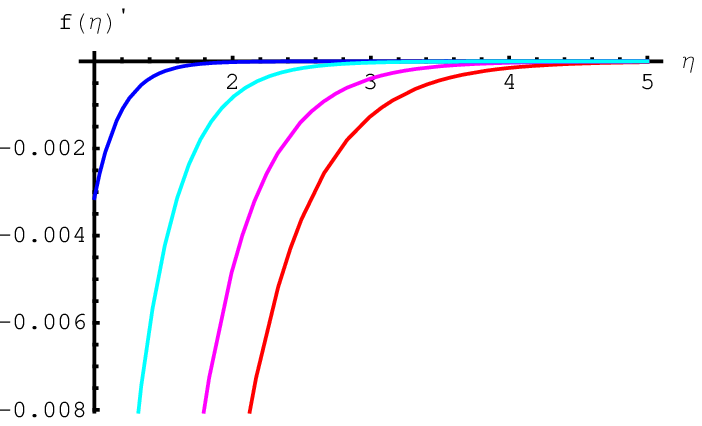, width=0.45\linewidth} \epsfig{file=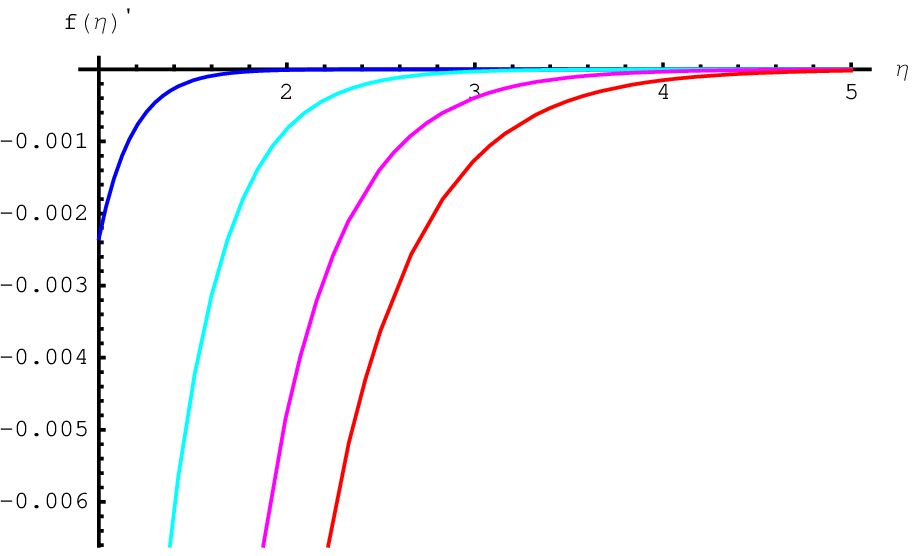,
width=0.45\linewidth}\caption{The left and right figures describe
the behavior of $f'(\eta)$ corresponding to lower and upper branch
solution, respectively for varying $\delta$ and $\gamma=0.1,~s=1,~M
=0.1.$ In both figures the blue, light green, pink and red curves
corresponds to $\delta =-0.01,~-0.02,~-0.03~-0.04,$ respectively.}
\end{figure}
\begin{figure}
\center\epsfig{file=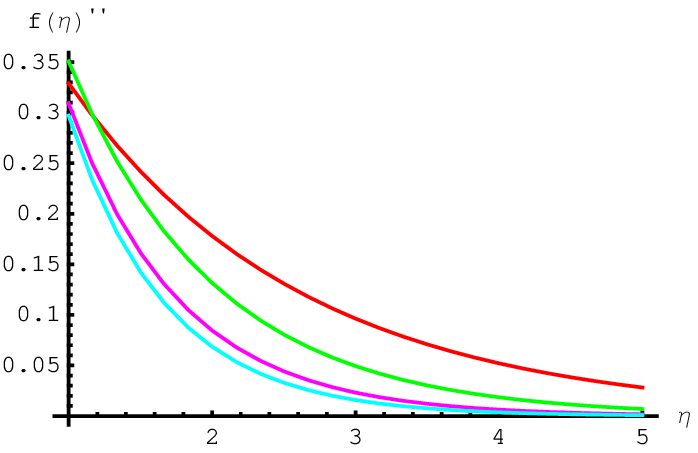, width=0.45\linewidth}
\epsfig{file=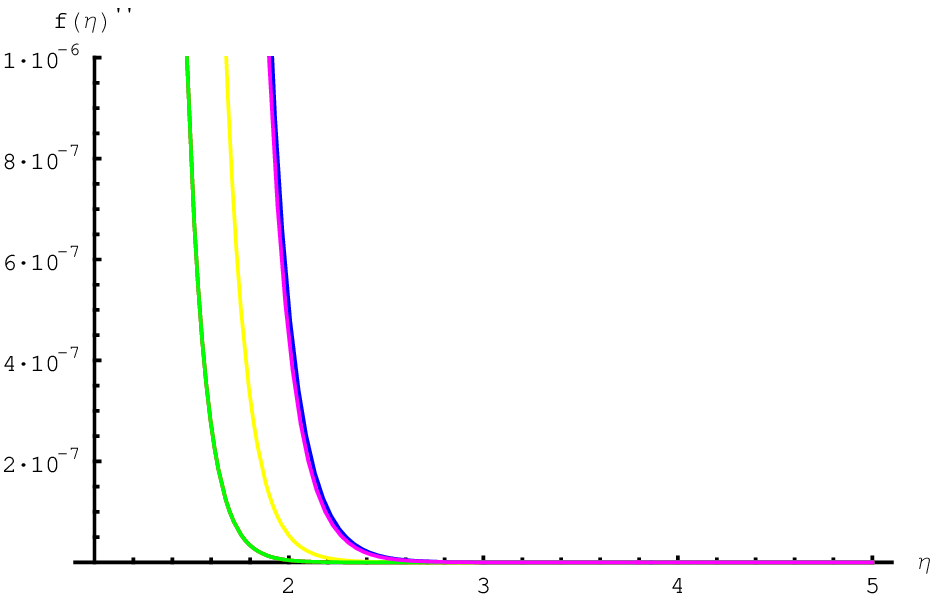, width=0.45\linewidth}\caption{The left and
right figures describe the behavior of $f''(\eta)$ corresponding to
lower and upper branch solutions, respectively for varying $\gamma$
and $s=1,~ \delta= -0.01,~M=0.1$. In the left figure light green,
pink, green and red curves correspond to
$\gamma=0.04,~0.02,~0.01,~0.1,$ respectively while in right ¯g- ure
blue, pink, brown and green curves correspond to
$\gamma=0.04,~0.02,~0.01,~0.1,$ respectively.}
\center\epsfig{file=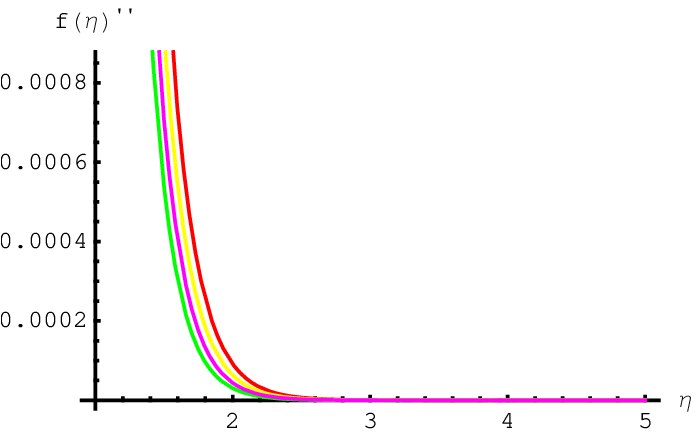, width=0.45\linewidth}
\epsfig{file=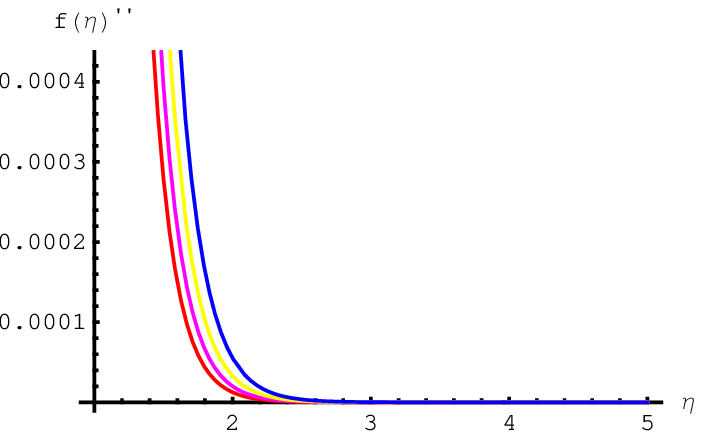, width=0.45\linewidth}\caption{The left and
right figures describe the behavior of $f''(\eta)$ corresponding to
lower and upper branch solutions, respectively for varying $s$ and
$\gamma= 0.1,~\delta =-0.01,~M = 0.1$. In the left figure red,
brown, pink and green curves corresponds to $s=0.1,~0.2,~0.3,~0.4,$
respectively while in right figure blue, brown, pink and red curves
correspond to $s = 0.1,~0.2,~0.3,~0.4,$ respectively.}
\end{figure}
\begin{figure}
\center\epsfig{file=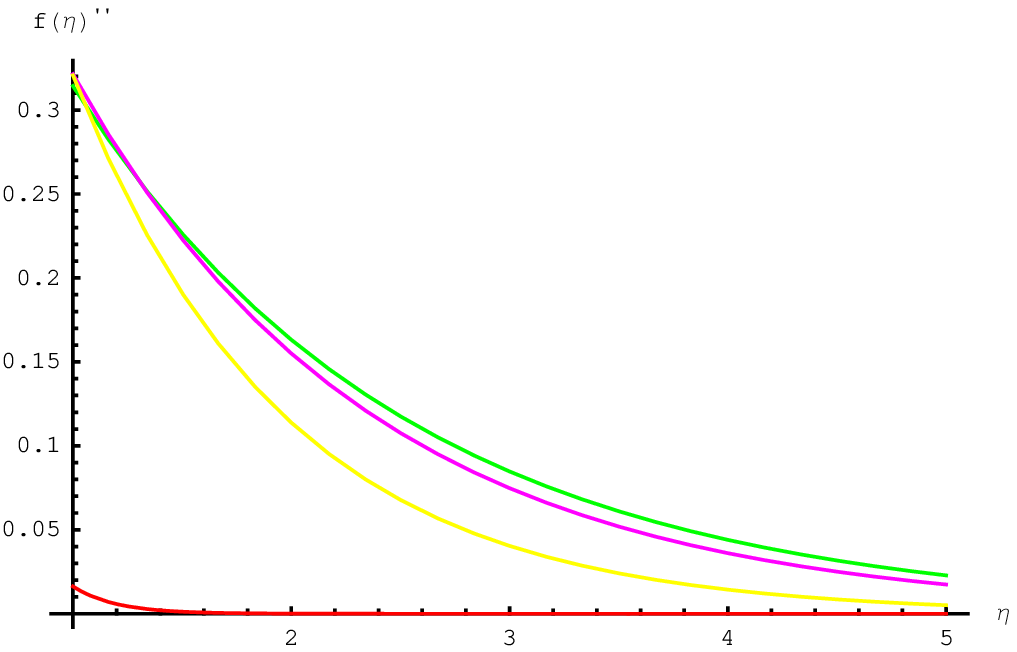, width=0.45\linewidth}
\epsfig{file=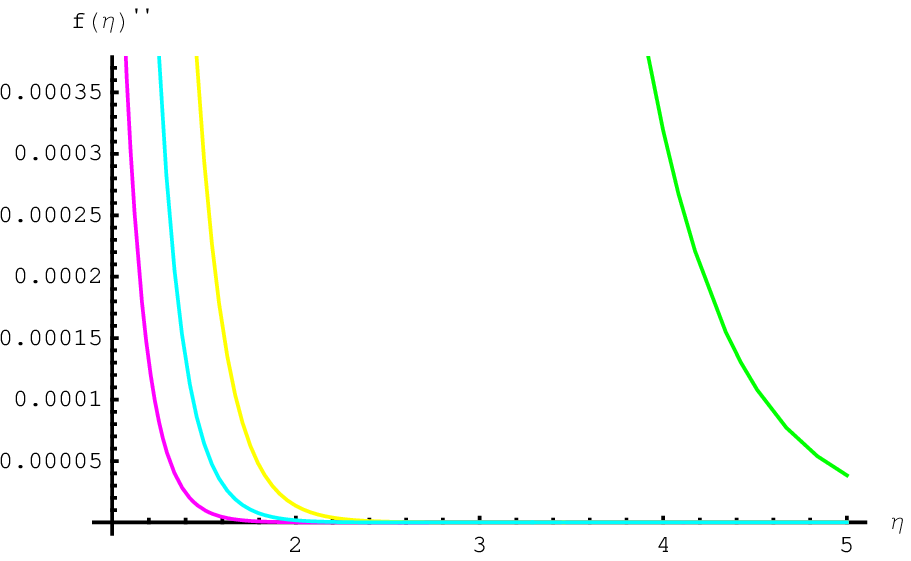, width=0.45\linewidth}\caption{The left and
right ¯gures describe the behavior of $f''(\eta)$ corresponding to
lower and upper branch solution, respectively for varying $M$ and $s
= 1$,$\delta =-0.01,~\gamma= 0.1$. In the left figure red, brown,
pink and green curves correspond to $M =0.1,~1,~ 2,~3,$ respectively
while in right figure green, brown, light green and pink curves
correspond to $M =0.1,~1,~ 2,~3,$ respectively.}
\center\epsfig{file=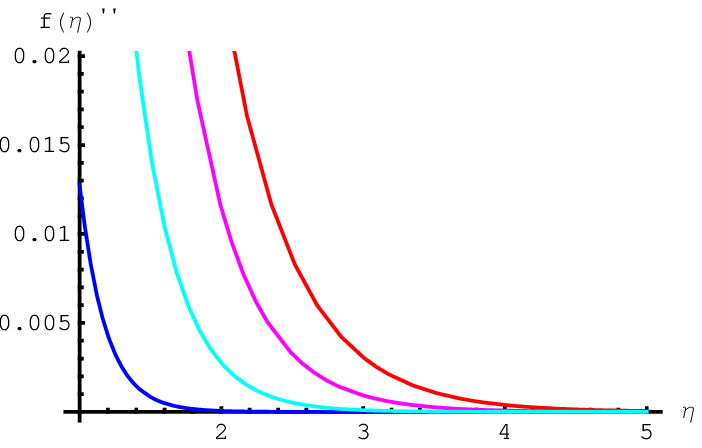, width=0.45\linewidth}
\epsfig{file=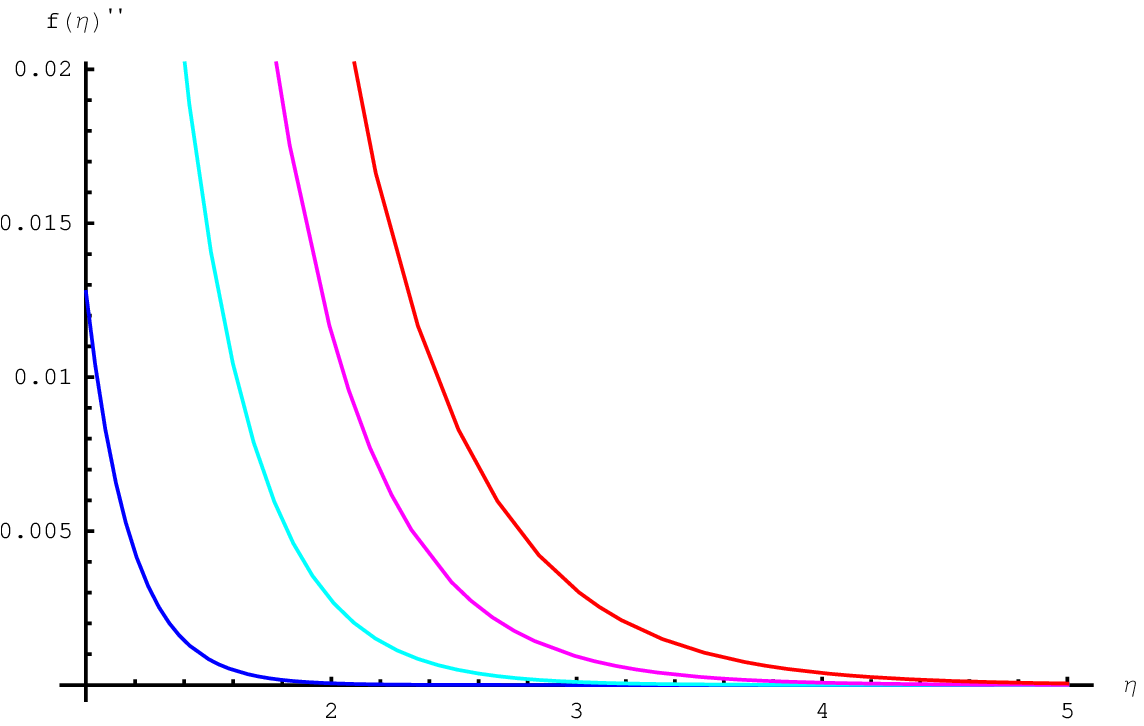, width=0.45\linewidth}\caption{The left and
right figures describe the behavior of  $f''(\eta)$ corresponding to
lower and upper branch solutions, respectively for varying $delta$
and $s=1,~M=0.1,~\gamma = 0.1$. In both figures the blue, light
green, pink and red curves corresponds to $\delta =
-0.01,-0.02,0.-03,-0.04,$ respectively.}
\end{figure}

The left graph in the figure \textbf{3} shows that for lower branch
solution the velocity of the shrinking sheet increase of $\delta$.
For maximum value of $\delta =-0.01$, velocity reduces from negative
finite values to zero. All the remaining curves imply that a sheet
moving with infinite velocity is highly damped due to
magnetohydrodynamical effects. The same behavior of the velocity
profile appear in case of upper branch solution with varying
$\delta$ is shown in right graph of figure \textbf{3}. In figure
\textbf{4} the lower and upper branch solution for the velocity of
the shrinking sheet have been plotted with varying $M$,
respectively. In left graph of figure \textbf{4} for small magnetic
effect velocity is always zero while all the others curves implies
that shrinking sheet moving with finite velocity either comes to
rest or moves with constant velocity as it becomes parallel to
$\eta$-axis.

The shear stress of the wall is shown in figures \textbf{5-8}. The
left graph in figure \textbf{5} shows that in case of lower branch
solution wall stress decreases with the increase of slip parameter
$\gamma$. The wall stress decreases as sheet moves with maximum
speed. Similar behavior of shear stress with increasing $\gamma$ is
presented in right graph of figure $\textbf{5}$ for upper branch
solution. All the stress profiles have same nature for the variation
of different parameters. From the both graphs in figure \textbf{7},
it can be deduced that shearing stress on the wall is linearly
related to the strength of magnetic field.

\section{Conclusion}

In this paper, we have investigated the exact solution of governing
Navier-Stokes equations for the magnetohydrodynamic viscous flow
over a shrinking sheet with second order slip. We have examined the
effect of slip parameters, magnetic parameters suction parameter on
the fluid flow. We have presented the velocity profile $f'(\eta)$
and wall stress profile $f''(\eta)$, for the different combination
$M,~\delta,~\gamma$ and $s$. It has been found that the effect of
$\delta< 0$ (due to shrinking sheet) on velocity and shear stress
profiles in upper and lower branch solutions case is same. An
interesting feature appear, when M is small stress is small and
velocity is zero. In case of varying $\delta$, velocity is highly
damped due to magnetic effects.

\end{document}